\newcommand{\rs}{\rho_s}

\newcommand{\gp}{\gamma}

\newcommand{\en}{\epsilon_n}
\newcommand{\ad}{\alpha_d}

\newcommand{\kp}{k_\parallel}

\newcommand{\pp}{\partial_\parallel}

\newcommand{\s}{\hat{s}}

\newcommand{\Dpe}{\nabla^2_\perp}
\newcommand{\lrb}{L_{0}}

\newcommand{\Lr}{L_r}
\newcommand{\Ly}{L_\theta}
\newcommand{\Om}{\Omega}
\newcommand{\Oc}{\omega_{\rm c}}
\newcommand{\Mc}{m_{\rm c}}

\documentstyle[twocolumn,aps]{revtex}
\input{psfig.sty}
\begin{document}
\draft
\author{K. Hallatschek}
\address{Max-Planck Institut f\"ur Plasmaphysik, EURATOM-IPP
Association, D-85748 Garching, Germany}
\title{Condensation of 
microturbulence-generated 
shear flows into global modes}
%\date{\today} 
\maketitle
\begin{abstract}
  In full flux-surface computer studies of tokamak edge turbulence, a
  spectrum of shear flows is found to control the turbulence level and
  not just the conventional (0,0)-mode flows. Flux tube domains too
  small for the large poloidal scale lengths of the continuous
  spectrum tend to overestimate the flows, and thus underestimate the
  transport. It is shown analytically and numerically that under
  certain conditions dominant (0,0)-mode flows independent of the
  domain size develop, essentially through Bose--Einstein condensation
  of the shear flows.
\end{abstract}
%body of paper here

The energy confinement of tokamaks is mainly controlled by small-scale
($\sim$cm) turbulence giving rise to the ``anomalous transport''.
Analytical and computer-aided studies have found that the anomalous
transport in turn is often controlled by fluctuating ``zonal flows''
\cite{diamondflow,burrellflow}, poloidal shear flows, which are
assumed to have zero poloidal and toroidal mode number, but have
radial scales similar to the turbulence. The present paper deals with
the question what happens to the shear flows when the turbulence
scale-lengths become very small compared to the plasma size such as in
the tokamak edge, in particular in future large machines.
In this limit, the shear flows can either contain a finite
$(0,0)$-mode component, or may loose their global character and change
into vortices with finite poloidal scale length, as will be
demonstrated by numerical full flux-surface edge turbulence studies.
Regarding their large poloidal and parallel but small (similar to the
turbulence) radial scale-lengths, these vortices should not be regarded
as drift-waves or convective cells \cite{convcell} but rather as
poloidally localized shear flows.

For cost reasons, the domains of turbulence simulations are usually
thin flux tubes \cite{beer} or tokamak sectors \cite{scott},
equivalent to flux tubes with special boundary conditions. The flux
tube dimensions perpendicular to the magnetic field are of the order
of $\sim10$cm and they extend several $\sim10$m along the magnetic
field to accomodate the prevalent turbulent structures. For poloidally
localized shear flows, however, these computational domains are not
adequate and the flows always appear to extend across the complete
flux tube. Since the $(0,0)$ mode is not damped as the other modes, it
may therefore exert a strong stabilizing effect on the turbulence,
which underestimates the transport compared to a true full
flux-surface simulation. For the core turbulence, which has relatively
large scales, full torus simulations \cite{lin,sydora} exhibit
zonal flows extending over the complete flux surfaces. However, even
for these scenarios it is not clear, whether this remains true for
much larger ratio of flux surface circumference to turbulence scale
length or whether the flows have a finite scale length in poloidal
direction.

Following the numerical results, an analytic model for the shear flows
is described, in which their poloidal and radial wavenumber spectra
are controlled by the interplay of damping by the collisional electron
response and ion dissipation, the linear response of the turbulence to
the flows, and the excitation of flows by random fluctuations. Under
certain conditions, a non-zero fraction of the flow energy is
generated as $(0,0)$-mode flows, {\em regardless of the system size}.
The mechanism is analogous to the Bose--Einstein condensation (BEC).
The three effects acting on the flows take the role of absorption,
stimulated, and spontaneous emission. For the BEC, a macroscopic
fraction of the quanta is eventually scattered into the ground state
because the state density {\em near} the ground state is too low to
hold sufficiently many quanta under the prevalent conditions. For the
flow system, the turbulence and shear flows form a feed back loop,
which regulates the shear flow energy to the level needed for
turbulence saturation. Condensation into the $(0,0)$ flow component
occurs when a threshold in required flow energy is exceeded, and the
$m\ne0$ modes are unable to hold it.

\paragraph*{Numerical results. ---}

We discuss the results of turbulence simulations of the three
dimensional electrostatic drift Braginskii equations with isothermal
electrons (a subset of the equations of Ref.~\cite{rogers}) for two
different cases: (a) the predominant instability is the resistive
ballooning mode with the nondimensional parameters $\ad=0.2$,
$\en=0.08$, $q=5$, $\tau=1$, $\eta_i=1$, $\s=1$; (b) there is a
significant contribution from ITG modes with $\ad=0.4$, $\eta_i=3$ and
the other parameters as in (a). The radial domain width in terms of
the resistive ballooning scale length, $\lrb$, in (a) was $24\lrb$ and
in (b) $48\lrb$, the width $\Ly$ perpendicular to ${\bf r}$ and ${\bf
  B}$ was $24\lrb$ [only for (a)], $192\lrb$, $384\lrb$, and $768\lrb$
(the corresponding tokamak minor radius is $a=\Ly q /(2\pi)$). For a
definition of these parameters and units see
Refs.~\cite{rogers,guzdar}. The parameters of the largest
domain are consistent with the physical parameters $R=3$~m, $a=1.5$~m,
$L_n=12$~cm, $q_0=3.2$, $n=3.5\times10^{19}$~m$^{-3}$, $Z_{\rm
  eff}=4$, $B_0=3.5$~T, and for (a) $T=100$~eV, $\lrb=5.1$~mm,
$\rs=0.58$~mm and for (b) $T=200$~eV, $\lrb=3.6$~mm, $\rs=0.82$~mm.
The perpendicular grid step size was $\Delta=0.19\lrb$ (a) and
$\Delta=0.38\lrb$ (b).  Parallel to the magnetic field 12 points per
poloidal connection length were sufficient due to the large parallel
scales of the ballooning modes. The largest runs had a grid of
$128\times4096\times12$.

\begin{figure}[p]
  \psfig{file=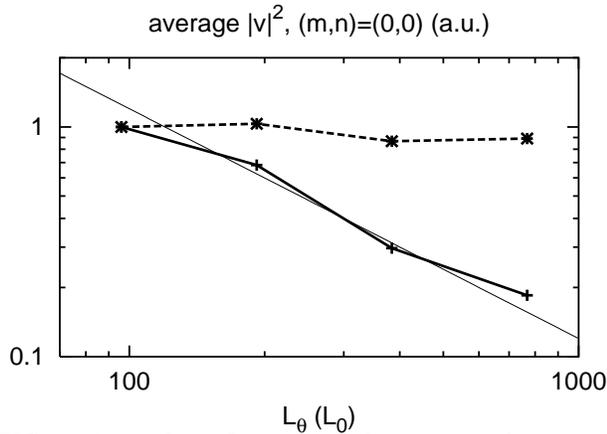,width=\linewidth}
  \caption{$(0,0)$ shear flow energy density as a function
    of poloidal domain size $\Ly$ for case (a) (solid) without
    condensation and case (b) (dashed) exhibiting condensation; the
    thin line is proportional to $1/\Ly$.}
  \label{fig:1}
\end{figure}

The dependence of the average $(0,0)$ shear flow energy density on the
domain size, $\Ly$, is compared for the two cases in Fig.~\ref{fig:1}.
In contrast to case (b), the shear flows in (a) are apparently not
condensed into the $(0,0)$ mode since its energy density decreases
proportional to $1/\Ly\propto1/a$, as is expected when a given shear
flow energy density is distributed equally among an increasingly dense
set of modes.

\begin{figure}[p]
  \psfig{file=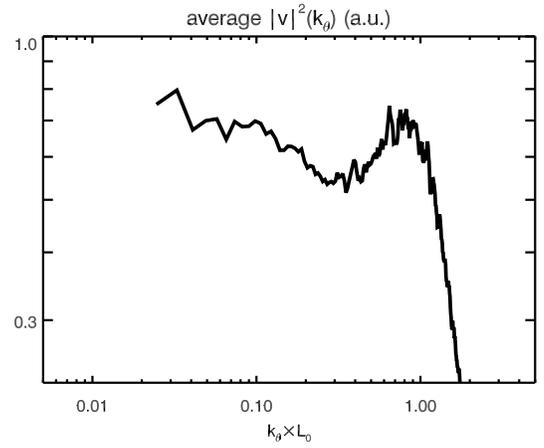,width=\linewidth}
  \caption{Mean square shear flow amplitude as a function of $k_{\theta}$ for 
    case (a) for $\Ly=768\lrb$. Note the greatly different scale
    lengths of the turbulence ($k_{\theta}>0.3$) and the shear
    flows ($k_{\theta}<0.3$).}
  \label{fig:2}
\end{figure}

The $k_\theta$ spectrum of the flow velocity,
$v=v_{\theta}=\partial_r\phi$, for the $\Ly=768\lrb$ runs [for case
(a) see Fig.~\ref{fig:2}] exhibits a rise at low $k_\theta$ associated
with the shear flows, different from the microturbulence fluctuations
at $k_\theta\sim 1$. The square amplitude of the $m=0$ mode in case
(a) and (b) is $0.3$ and $7$ times, respectively, the total shear flow
amplitude, suggesting strong condensation for (b). In both cases, the
typical poloidal scale length of the $m\ne0$ shear flows is roughly a
factor $10$ greater than the scales of the turbulence. Failure of the
computational domain to accomodate the scales of the uncondensed shear
flows in case (a) results in an overestimate of the shear flow
amplitude, and hence in an underestimate of the anomalous transport.
The particle flux for $\Ly=\Lr=24\lrb$ was found to be $25\%$ lower
than for $\Ly=768\lrb$.

\paragraph*{Analytic model. ---}

As the first ingredient of a qualitative model for the poloidal shear
flow spectra, we calculate the linear dispersion relation for finitely
elongated shear flows. For clarity, in the linear electrostatic
vorticity equation (with the plasma parameters absorbed into the
units, see, e.g., \cite{guzdar}),
\begin{equation}
  \Dpe (\partial_t+\gp)\phi+\pp^2 \phi=0,
\label{eq:fleq}
\end{equation}
we neglect temperature fluctuations, parallel ion velocity, drift
effects, curvature and magnetic fluctuations. These effects can lead
to a real frequency (e.g., geodesic acoustic modes \cite{gam,gam2})
and to a coupling to parallel sound waves or Alfv\'{e}n waves. The
dissipative effects are the flow damping $\gp$ due to the ion
dissipation assumed independent of the wavenumber and the damping due
to the resistive electron response. As we will see below, for a
potential condensation of the shear flows into global modes only a
small region around a certain radial wavenumber, $k_0$, is important,
which we set to one in (\ref{eq:fleq}) since its absolute value is not
important.  Because of the large poloidal wavelengths of the flows, we
approximate $-\Dpe\approx k_0^2=1$ and obtain the dispersion relation 
\begin{equation}
\omega_{{\rm lin}}=-i\left(\gp+\kp^2\right),\qquad
\kp=\left(\frac{m}{q(r)}-n\right).\label{eq:lineq}
\end{equation}
The damping by the parallel resistive electron response is weak if
either $m=n=0$ holds, or $r$ is near a resonant surface defined by
$m-n q(r_{mn})=0$.  Focusing on a thin region around $r=r_0$ we obtain
$\kp\approx m\alpha_0(r-r_{mn})$, $\alpha_0=-q'(r_0)/q(r_0)^2$. Hence
the resistive flow damping is proportional to $m^2$, which is the
reason for the poloidal elongation of the flows, i.e., their low mode
numbers.

As reaction to a shear flow
\cite{diamondflow,burrellflow,flowcurv,fluxresponse} the
microturbulence may in turn influence the flows via the Reynolds
stress \cite{negvisc,gruzinov} or the Stringer--Winsor mechanism due
to poloidal pressure asymmetries \cite{gam,hassam}. Restricting
ourselves to linear response theory, we assume a (coherent) flow
amplification rate $g(k_r)$ depending only on the radial wavenumber
$k_r$, because of the large poloidal correlation lengths of the shear
flows.
With the (incoherent) random forcing, $f$, representing the effect of
the turbulence fluctuations, the equation for the flow amplitude in
frequency space has the form of a Langevin equation,
\begin{equation}
\partial_t {v}=-i\omega_{{\rm lin}}{v}
+g(k_r){v}+{f}.\label{eq:fullflow}
\end{equation}

From (\ref{eq:fullflow}) we obtain a relation between the mean square
spectra of the flows and the forcing in frequency space,
\begin{equation}
{\overline{|\hat{v}|^2}}=\frac{{\overline{|\hat{f}|^2}}}{\omega^2+
  (-i\omega_{{\rm lin}}+g(k_r))^2}.\label{eq:eqeq}
\end{equation}
Assuming that ${\overline{|\hat{f}|^2}}$ is independent of ${\bf
  k},\omega$ (white noise), the integration of (\ref{eq:eqeq}) over
$\omega$ yields the relation between the mean square flow amplitude at
an instant of time and the forcing,
\begin{equation}
\overline{|v|^2}=\frac{\overline{|\hat{f}|^2} \pi}{|-i\omega_{{\rm
      lin}}+g(k_r)|}.
\label{eq:flow}
\end{equation}
The flow intensity (\ref{eq:flow}) replaces the Bose distribution in
the BEC case. Both functions tend to infinity when the amplification
(stimulated emission) terms approach the damping (absorption) terms.
As long as every mode is net-damped at a rate independent of the
system size, the energy density stored in (0,0) modes must decrease
proportional to the system size, since the total shear flow energy is
distributed among an increasingly dense set of modes. However,
analogous to the thermodynamic theory of the BEC, when the continuous
flow spectrum is unable to hold the shear flow energy for non-zero
minimum net-damping rate and given random forcing, the nonlinear flow
amplification term must adjust so that the remaining part of the flow
energy is excited in the form of the most weakly damped modes, which
are (0,0) modes.
Hence, to demonstrate the possibility of condensation, it has to be
shown that the flow amplitude in $m\ne0$ modes stays finite when the
net-damping rate of the $m=0$ modes tends to zero, in the limit of
infinite system size or, equivalently, in the approximation of a
continuous poloidal mode spectrum.

It is sufficiently general, to assume that $g(k_r)$ has a maximum at
$k_r=k_0$ of order of the turbulence wavenumbers and is parabolic near
that maximum, $g(k_r)=g_0-g_1(k_r-k_0)^2$. The amplification terms
will nearly cancel the damping terms only for wavenumbers near $k_0$,
which justifies the approximation $k_r\approx k_0$ which was made in
the derivation of (\ref{eq:lineq}). For the following analysis we
shift the $k_r$ spectrum of the flows so that $k_0=0$.  With $i
k_r=\partial_r$, the operator in the denominator of (\ref{eq:flow}),
\begin{equation}
(\gp-g_0)-g_1 \partial_r^2+ \left(m \alpha_0 (r-r_{mn}) \right)^2,
\end{equation}
is the quantum mechanical Hamiltonian of the harmonic oscillator. 
Its eigenvalue for a mode with the ``quantum numbers''
$(m,r_{mn},l)$, $l\in\{0,1,2,\dots\}$, is
\[
\omega_l=\gp-g_0 + 2\sqrt{g_1 } \left|\alpha_0 m\right| (l+1/2).
\]
The sum over $l$ of the eigenmode contributions to (\ref{eq:flow}) at
fixed $(m,r_{mn})$ results in a logarithmic divergence, which stems
from the infinitely broad random forcing spectrum and infinitely fast
turbulence response. Hence, we cut off $\omega_l$ at an appropriate
$\Oc$ depending on the turbulence.
The sum is then approximated by an integral over $l$. The resulting
amplitude associated with each pair $(m,r_{mn})$ is
\begin{equation}
\overline{|v|^2}(m,r_{mn})=\frac{\overline{|\hat{f}|^2}\pi}{2\sqrt{g_1}
|\alpha_0m| } \ln\frac{\Oc}{\omega_0}
\label{eq:resampl}
\end{equation}
with $\omega_0=\gp-g_0 + \sqrt{g_1} |\alpha_0m|<\Oc$. The density of
rational surfaces is $|\alpha_0 m|$ for given $m$. Approximating the
sum over all $\overline{|v|^2}(m,r_{mn})$ contributions to
(\ref{eq:flow}) with $m\ne0$ by an integral (which becomes exact for
infinite system size), the total instantaneous energy density of the
flow modes with $m\ne0$,
\begin{equation}
\overline{|v|^2}_{m\ne0}=\int_{-\Mc}^{\Mc} \left|\alpha_0
 m\right|  \cdot\overline{|v|^2}(m,r_{mn}) dm , \label{eq:flow2}
\end{equation}
is obtained, where the integration interval is limited by the cutoff
$\Mc$ defined by $\omega_0(m=\Mc)=\Oc$. With the minimum net-damping
rate $\Om=\omega_0(m=0)=\gp-g_0$ we obtain
\begin{equation}
\overline{|v|^2}_{m\ne0}=
\frac{\pi\overline{|\hat{f}|^2}}{2 |\alpha_0| g_1}
\left[\Oc-\Om
\left(1+\ln\frac{\Oc}{\Om}\right)\right].
\label{eq:spaceampl}
\end{equation}
This expression converges to a finite value for $\Om\rightarrow0$. On
the other hand, because the energy density of the $m=0$ modes, which
tends to infinity for $\Om\rightarrow0$ [the integral over $k_r$ of
(\ref{eq:flow}) does not exist for $-i\omega_{{\rm lin}}+g_0=0$], has to be
finite, we always have $\Om>0$. If the turbulence saturation requires
a higher flow level than (\ref{eq:spaceampl}) at $\Om\rightarrow0$,
the description of the system by a continuum of poloidal mode numbers
breaks down, the flow energy which can not be received by the $m\ne0$
modes condenses into $m=0$ modes, and simultaneously
$\Om\rightarrow0$.

In a similar manner, it can be shown that in the limit of infinite
system size the $m=0$ condensate is {\em completely} contained in the
$n=0$ modes. Furthermore, there is no condensation of the radial wave
numbers but the $k_r$ spectrum becomes arbitarily narrow around the
point of weakest net-damping for large system size.

\begin{figure}[p]
  \psfig{file=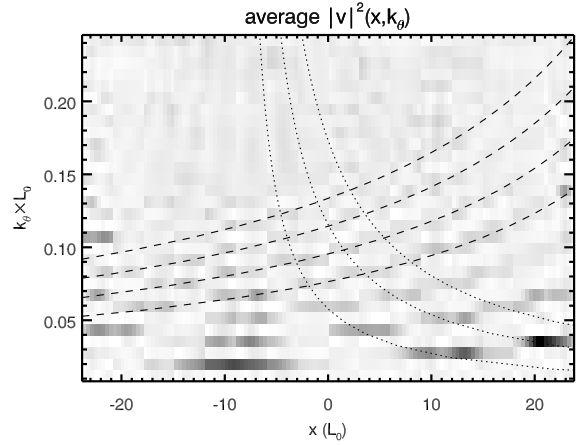,width=\linewidth}
  \caption{Mean square shear flow amplitude as a function of $k_{\theta}$ and 
    radius $x=r-r_0$ for case (b) for $\Ly=768\lrb$. Note the
    localization of $k_{\theta}\ne0$ flows on resonant surfaces. The
    resonances lie on the intersections of families of hyperbolae in
    the $x\times k_{\theta}$ plane, some examples of which are
    displayed.}
  \label{fig:3}
\end{figure}

\begin{figure}[p]
  \psfig{file=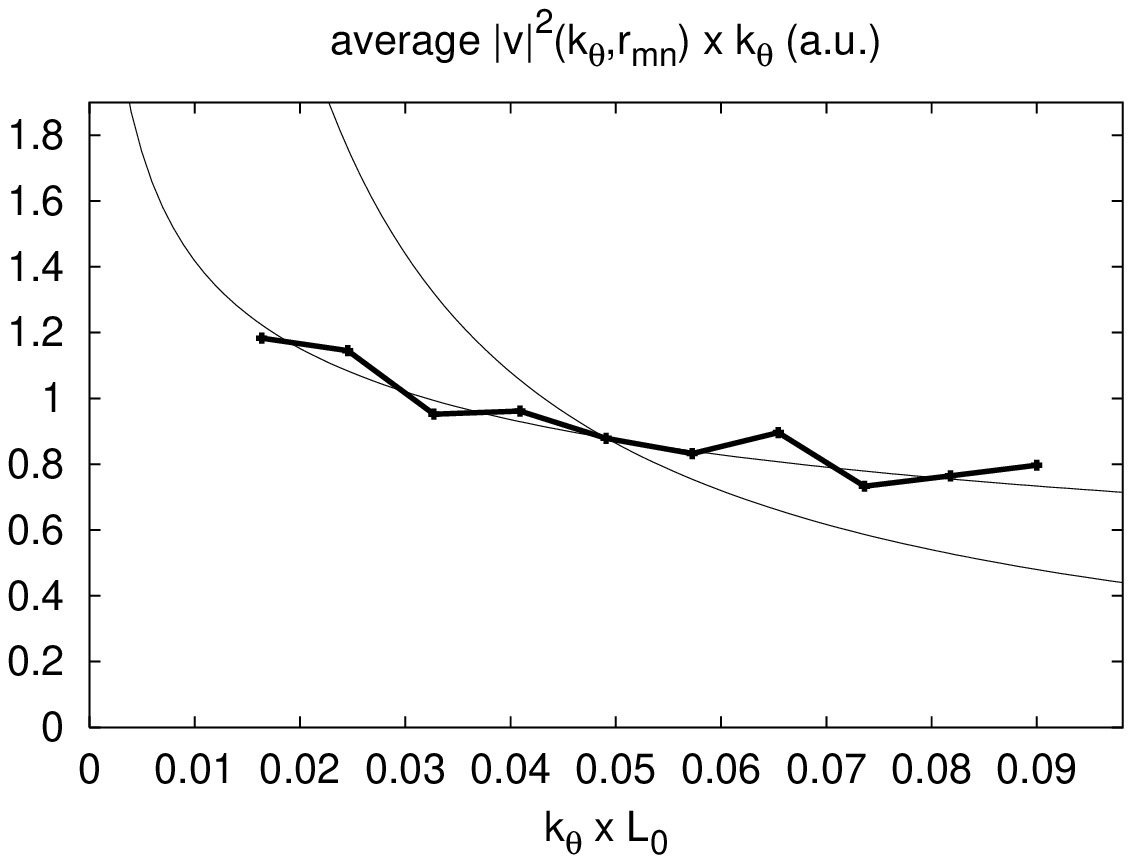,width=\linewidth}
  \caption{Mean square shear flow amplitude on a single resonant
    surface multiplied with $k_{\theta}\propto m$ for case (b) for
    $\Ly=768\lrb$. The thin lines are proportional to
    $k_{\theta}^{-1}$ (steeper curve) and $k_{\theta}^{-0.3}$ (flatter
    curve).}
  \label{fig:4}
\end{figure}

Strictly speaking, condensation is unprovable by numerical studies,
due to the restriction to finite system sizes. However, the validity
of the individual parts of the model can be checked in the
simulations. The localization of the $m\ne0$ shear flows on resonant
surfaces is obvious in a plot of the flow spectra versus radius
(Fig.~\ref{fig:3}). The total flow amplitude associated with each
$(m,r_{mn})$ quantum number in Eq.~(\ref{eq:resampl})
(Fig.~\ref{fig:4}) has much weaker slope than $k_\theta^{-2}\propto
m^{-2}$ for $k_\theta,m\rightarrow0$. Therefore the integral over all
$m\ne0$ shear flows (\ref{eq:flow2}) is expected to be finite in the
limit of infinite system size (even if the estimate (\ref{eq:resampl})
for the individual amplitudes should be quantitatively wrong).
Furthermore, the integral is reasonably well approximated by the
corresponding sum in the finite system. Consequently, the infinite
system will have approximately the same ratio of $m=0$ flow amplitude
to $m\ne0$ flow amplitude.  Finally, we note that the numerical
studies agree with the above analytical prediction, that the flow
condensate exhibits a strong peaking in $k_r$ for sufficiently large
system size.

\paragraph*{Conclusions and consequences. ---}
It has been shown numerically that in general the shear flows
controlling the turbulence are not only $(0,0)$ modes but rather
consist of a spectrum of poloidal mode numbers. The $(m,n)\ne(0,0)$
flows differ from drift waves or convective cells by their large
poloidal [10 times larger than the turbulence (Fig.~\ref{fig:2})] and
parallel scale length, while their perpendicular scale length is
similar to that of the turbulence.  These shear flows are localized in
the vicinity of resonant surfaces (Fig.~\ref{fig:3}). In the limit of
large system size, a non-zero (0,0)-mode amplitude develops only if
the shear flows undergo a condensation into these modes, analogous to
the Bose--Einstein condensation. Several features predicted by the
analytic model have been reproduced by the numerical simulations.

Due to the cancelling of damping and amplification terms, the $(0,0)$
flow condensate is practically undamped. This means in quantum
mechanical language that the rate of absorption and incoherent
re-emission, the ``collision'' rate, vanishes. Hence, far ranging
interactions or ordering effects might be mediated via the shear flow
condensate (but not by the uncondensed flows that suffer collisions
and are pinned to resonant surfaces). As a consequence in the simple
system used here the $k_r$ spectra become arbitrarily narrow.

Since the flows depend on the distribution of rational surfaces and
mode numbers, to accurately model the shear flow system in numerical
studies, care has to be taken to not introduce spurious resonant
surfaces or modes, e.g., by parallel extension of the flux tube
\cite{beer,scott}. Remarkably, it can be shown that increasing
the flux tube length does not lead to the correct limit of large
system sizes, since, e.g., for an infinitely long flux tube,
condensation into $(0,0)$ modes can not occur.

Up to now, in flux tube based turbulence computations the shear flows
were implicitly assumed to be global modes. With domain widths too
small for the large poloidal scales of the continuous part of the flow
spectrum, the flows {\em appear} to have zero poloidal and toroidal
mode number. Such modes do not experience the resistive damping, which
would reduce the flow amplitude in a full system.  Hence, the
simulations tend to overestimate the total flow amplitude, which may
therefore exert a strong stabilizing effect on the turbulence. To
avoid an underestimate of the transport, flux tube simulations have to
be checked for influences of a finite poloidal scale length of the
flows.

The author would like to thank Dr. D. Biskamp for valuable
discussions.

\end{document}